\documentclass[aps,pre,twocolumn,groupedaddress]{revtex4-2}
\usepackage{amsmath,amssymb}
\usepackage{slashed}
\usepackage{graphicx}
\usepackage{hyperref}
\usepackage{cleveref}
\usepackage[dvipsnames]{xcolor}

\newcommand{\dd}{\text{d}}
\renewcommand{\vec}[1]{\boldsymbol{#1}} % use boldsymbol for greek letters
\newcommand{\mat}[1]{\underline{#1}}
\newcommand{\mmat}[1]{\underline{\underline{#1}}}
\newcommand{\scalar}[2]{\left(#1,#2\right)}
\begin{document}

%\preprint{}

\title{Generalized Langevin dynamics simulation with non-stationary memory kernels:\\ How to make noise}


\author{Christoph Widder}
\author{Fabian Glatzel}
\author{Tanja Schilling}
\email[]{Tanja.Schilling@physik.uni-freiburg.de}
%\thanks{}
\affiliation{Institut f\"ur Physik, Albert-Ludwigs-Universit\"at Freiburg,\\ Hermann-Herder-Stra\ss e 3, 79104 Freiburg im Breisgau, Germany}

\date{\today}

\begin{abstract}
We present a numerical method to produce stochastic dynamics according to the generalized Langevin equation with a non-stationary memory kernel. This type of dynamics occurs when a microscopic system with an explicitly time-dependent Liouvillian is coarse-grained by means of a projection operator formalism. We show how to replace the deterministic fluctuating force in the generalized Langevin equation by a stochastic process, such that the distributions of the observables are reproduced up to moments of a given order. Thus, in combination with a method to extract the memory kernel from simulation data of the underlying microscopic model, the method introduced here allows to construct and simulate a coarse-grained model for a driven process. 
\end{abstract}

\maketitle

\section{Introduction}\label{sec:introduction}
In computational biology, fluid-dynamics engineering and soft matter physics the dynamics of complex systems is often modeled by means of the Langevin equation \cite{snook2006,zwanzig2001nonequilibrium,parish2017}. The basic idea of this approach is to simulate a set of "relevant" degrees of freedom while the effect of the remaining degrees of freedom is treated by means of friction terms and stochastic forces.
Projection operator formalisms are a set of methods that allow to derive these effective equations of motion, in principle, rigorously from the underlying microscopic dynamics \cite{Mori_65,Zwanzig_61,grabert2006projection,Vrugt2020,Hijon_2010,chorin:2000,ottinger2002}. After having been a very active research field from the 1960s to the 1980s, followed by three decades of somewhat lesser research activity, recently the development of models and of computer simulation methods by means of projection operator formalisms has regained much in popularity again \cite{espanol2009,MeyerVoigtmann_2017,Charbonneau2018,Krommes2018,vogel2020,izvekov2021,ayaz2021,jung2022,vroylandt2022,Ayaz2022,vrugt2022}. In particular, the use of time-dependent projection operators to model the dynamics of systems which are far from equilibrium has moved into the focus of the soft matter modeling community \cite{schilling2022,MeyerVoigtmann_2019,Izvekov,Kawai_2011,vrugt2019}.

In this article we discuss the the non-stationary, linear, generalized Langevin equation, i.e.~the type of Langevin equation which one obtains when applying a linear ("Mori-type"), time-dependent projection operator to microscopic dynamics with an explicitly time-dependent Liouvillian \cite{MeyerVoigtmann_2019}. In previous work, we have shown how to extract the non-stationary memory kernel of this equation from experimental data  or data from computer simulations \cite{Meyer_2020,Meyer_Wolf}. Here we will show how to replace the fluctuating force by a set of random processes such that the dynamics generated by the resulting stochastic Langevin equation corresponds to the dynamics of the deterministic non-stationary, linear, generalized Langevin on the level of moments of a given order. I.e.~we show how to construct a coarse-grained model for a system out of equilibrium that reproduces the dynamics of a given set of observables \footnote{We will not discuss the case of a stationary memory kernel. Numerical methods to tackle this problem are presented e.g.~in \cite{leimkuhler2022}}.

For the case in which the memory kernel can be expressed in terms of a sum of damped oscillations, Wang et al.~recently presented a method to generate suitable noise \cite{wang2021}. The method we present here addresses a more general case.

\section{Generalized Langevin equation}\label{sec:gle}
We briefly recall how to derive the generalized Langevin equation from microscopic dynamics, which are governed by a time-dependent Liouvillian \cite{Zwanzig_61, Mori_65,grabert2006projection,MeyerVoigtmann_2017,MeyerVoigtmann_2019,schilling2022}. As an example, the reader could imagine a polymer melt under external mechanical driving, for which one would like to construct a coarse-grained model based on results from an atomistic simulation. Here, we show the case of real-valued, vectorial observables (e.g.~the positions of united atoms in the case of the polymer melt). The complex-valued case is discussed in appendix \ref{sec:complex_observables}.

We introduce a phase space observable $\vec{A}:\mathbb{R}^{d}\ni\Gamma\mapsto\vec{A}(\Gamma)\in\mathbb{R}^n$ and a phase space trajectory $\gamma(t;t_0,\Gamma_{t_0})$ with $\gamma(t_0;t_0,\Gamma_{t_0})=\Gamma_{t_0}$, where $\Gamma$ is some point in phase space and the parameters on the right-hand side of the semicolon denote the initial values of the trajectory. (Note that all derivations and results in this article hold true for explicitly time-dependent observables, too. This can be seen using the concept of an augmented phase space and following the line of arguments presented in ref.~\cite{MeyerVoigtmann_2019}.) The stream lines are denoted by $\dot{\Gamma}(t,\Gamma)$, where $\dot{\gamma}(t;t_0,\Gamma_{t_0})=\dot{\Gamma}(t,\gamma(t;t_0,\Gamma_{t_0}))$. Further, the time evolution operator $\mathcal{U}(t,t_0)$ is introduced as
\begin{equation}\label{eq:time_op}
	\vec{A}_t:=\mathcal{U}(t,t_0)\vec{A}:= \vec{A}(\gamma(t;t_0,.)) \, .
\end{equation}
The formal solution of the time evolution operator is obtained by taking the time derivative of this definition and using the fact that the stream lines are a function of time and phase space. Thus
\begin{equation*}
\mathcal{U}(t,t_0) := \exp_-\left( \int^t_{t_0}\dd\tau \mathcal{L}(\tau) \right) \, ,
\end{equation*}
where we introduced the Liouville operator $\mathcal{L}(t):=\dot{\Gamma}(t,.)\cdot \partial_\Gamma$ and the negatively time-ordered exponential $\exp_-$, see appendix \ref{app:gle1}.

Further, we introduce the time-dependent cross-correlation matrix $(\vec{X},\vec{Y})_t$, the Mori projection operator $\mathcal{P}(t)$, and its orthogonal complement $\mathcal{Q}(t)$
\begin{align}
	\scalar{\vec{X}}{\vec{Y}}_t &:= \int \dd\Gamma \, \rho(t,\Gamma) \vec{X}(\Gamma)  \vec{Y}^\top(\Gamma) \, , \\
	\mathcal{P}(t)\vec{X}&:=\scalar{\vec{X}}{\vec{A}}_t \cdot \scalar{\vec{A}}{\vec{A}}_t^{-1}\cdot \vec{A}\label{eq:MoriProjector} \, , \\
	\mathcal{Q}(t)&:=1-\mathcal{P}(t) \, ,
\end{align}
where $\rho(t,\Gamma)$ denotes the phase space density. Note that the ordering of the matrices and arguments of the cross-correlation in \cref{eq:MoriProjector} is uniquely fixed by the idempotence of $\mathcal{P}(t)$ as opposed to the one-dimensional case, refs.~\cite{MeyerVoigtmann_2017,
MeyerVoigtmann_2019}.
Further, $\scalar{\vec{A}}{\vec{A}}_t$ in \cref{eq:MoriProjector} must be invertible.
The cross-correlation matrix satisfies
\begin{equation*}
    	\scalar{\vec{X}}{\vec{Y}}_t= \scalar{\mathcal{U}(t,t')\vec{X}}{\mathcal{U}(t,t')\vec{Y}}_{t'} \, ,
\end{equation*}
and is reduced to a genuine scalar product in the one-dimensional case. 

The equation of motion for the trajectory $\vec{A}_t$ is obtained by applying the Dyson-Duhamel identity to the orthogonal contribution of the observable. The final result is the \textit{linear non-stationary generalized Langevin equation} (nsGLE)
\begin{align}
	\vec{\dot{A}}_t &=\mat{\omega}(t)\vec{A}_t + \int^t_{{t_0}}\dd s\,\mat{K}(t,s) \vec{A}_s \, +  \vec{\eta}_{t{t_0}}\, ,  \label{eq:gle2}
\end{align}
where the \textit{drift} $\mat{\omega}(t)$, the \textit{fluctuating forces} $\vec{\eta}_{ts}$ and the \textit{memory kernel} $\mat{K}(t,s)$ are defined by 
\begin{align}
	\mat{\omega}(t) &:= \scalar{\mathcal{L}(t)\vec{A}}{\vec{A}}_t\scalar{\vec{A}}{\vec{A}}^{-1}_t \, , \label{eq:drift}\\
	\mat{K}(t,s) &:= -\scalar{\vec{\eta}_{ts}}{\vec{\eta}_{ss}}_s\scalar{\vec{A}}{\vec{A}}^{-1}_s 
	  \label{eq:definitionMemoryKernel}\, ,\\
	\vec{\eta}_{ts} &:= \mathcal{Q}(s)\mathcal{G}_-(t,s)\mathcal{L}(t)\vec{A} \, , \label{eq:ff1}\\
	 	\mathcal{G}_-(t,s) &:= \exp_-\left(\int^t_{s}\dd\tau \mathcal{L}(\tau)\mathcal{Q}(\tau) \right) \, ,\label{eq:ff2}
\end{align}
see appendix \ref{app:gle} for details. 
By construction, the cross-correlation between the fluctuating force and the observable vanishes, i.e.
\begin{equation}
	\scalar{\vec{\eta}_{tt'}}{\vec{A}}_{t'} = \mat{0} \, . \label{eq:mean_f_A}
\end{equation}
With this, the memory kernel may be written as  
\begin{equation}\label{eq:fdt}
	\mat{K}(t,s) = -\scalar{\vec{\eta}_{tt'}}{\vec{\eta}_{st'}}_{t'}\scalar{\vec{A}}{\vec{A}}^{-1}_s \, , 
\end{equation}
where $t'$ may be chosen arbitrarily. This relation is known as the generalized or second \textit{fluctuation-dissipation theorem}\cite{grabert2006projection}.

Finally, we obtain the equation of motion for the auto-correlation function $C(t,t')$ by inserting eq.~(\ref{eq:gle2}) into the time derivative $\frac{\dd}{\dd t} C(t,t')$. Conveniently, we may choose the initial time to be $t_0:=t'$ in order to eliminate the fluctuating forces by applying eq.~(\ref{eq:mean_f_A}).
\begin{align}
	\mat{C}(t,t')&:=\scalar{\vec{A}_t}{\vec{A}_{t'}}_{t_0}  \, ,\label{eq:def_two_time_correlation} \\
	\frac{\dd}{\dd t}\mat{C}(t,t') &= \mat{\omega}(t)\mat{C}(t,t') + \int^t_{t'}\dd s\, \mat{K}(t,s) \mat{C}(s,t') \, . \label{eq:eom_ac}
\end{align}

%%%%%%%%%%%%%%%%%%%%%%%%%%%%%%%%%%%%%%%%%%%%%%%%%%%%%%%%%%%%%%%%%%%%%%%%%%%%%%%%%%

Nordholm and Zwanzig pointed out in ref.~\cite{nordholm1975} that powers of the same observable are an interesting exemplary set of observables for \cref{eq:gle2}. Let $X$ denote the observable of interest and choose $\vec{A}=\left(X,X^2,\cdots,X^m\right)^\top$ with $m\in\mathbb{N}$. Then the nsGLE for the first component reads
\begin{align}
  \label{eq:nonlinear}
	\frac{\dd X_t}{\dd t} &= \sum\limits_{i=1}^m\left(\mat{\omega}_{1,i}(t)X_t^i +\int\limits_{t'}^t\dd\tau\,\mat{K}_{1,i}(t,\tau)X_\tau^i\right) + \left(\boldsymbol{\vec{\eta}}_{1}\right)_{t't}.
\end{align}
Hence, one obtains an nsGLE containing drift and memory terms, which are non-linear in the observable. The terms may be regarded as a Taylor expansion in the observable and, thus, may provide a way to connect the exact equation of motion, \cref{eq:nonlinear}, to Landau Ginzburg models \cite{chaikin1995}, phase field models \cite{steinbach2009} and related approximative approaches that use Langevin dynamics in free energy landscapes, which are expressed as a series expansion in powers of the order parameter. We give a detailed discussion about equations of motions with a potential of mean force and (non-)linear memory in the observable in ref.~\cite{Glatzel_2021}. Including a constant into the set of observables one projects on, e.g. $\vec{A}=(1,X)^\top$, leads to a vanishing mean of the fluctuating force $\langle\vec{\eta}_{ts}\rangle\equiv0$. However, it will also cause an additional term in the nsGLE which depends on time only \cite{vroylandt2022,Kauzlaric2011}. Depending on the specific system and observable(s) one intends to describe, this might or might not be desirable. Here, we will not elaborate these special cases further and continue with the derivation of general relations between the terms in the nsGLE.\\
%%%%%%%%%%%%%%%%%%%%%%%%%%%%%%%%%%%%%%%%%%%%%%%%%%%%%%%%%%%%%%%%%%%%%%%%%%%%%%%%%%

\section{Calculation of the Memory Kernel}
Differentiating \cref{eq:eom_ac} with respect to $t'$ yields a Volterra integral equation for the memory kernel $\mat{K}(t,.)$ for each time $t$,
\begin{align*}
	&\mat{K}(t,t') =  \int^t_{t'}  \mat{K}(t,s)\scalar{\vec{A}_s}{\vec{\dot{A}}_{t'}}_{t_0}  \scalar{\vec{A}}{\vec{A}}^{-1}_{t'} \dd s \, + \\
	&+\mat{\omega}(t)\scalar{\vec{A}_t}{\vec{\dot{A}}_{t'}}_{t_0} \scalar{\vec{A}}{\vec{A}}^{-1}_{t'} -\scalar{\vec{\dot{A}}_t}{\vec{\dot{A}}_{t'}}_{t_0}  \scalar{\vec{A}}{\vec{A}}^{-1}_{t'}  \, .
\end{align*}
For continuous coefficient functions, there exists a unique continuous solution which can be solved by a Picard iteration \cite{burton}. Here, the numerical calculations of the memory kernel follow the algorithm described in ref.~\cite{Meyer_2020,Meyer_Wolf}. However, in these references, the algorithm is presented for a single, scalar-valued observable only, thus, the following adaptations to the case of vectors should be noted:\\
First, as mentioned in \cref{sec:gle}, the order of the individual terms in the projector (\cref{eq:MoriProjector}) is interchanged to get a proper projector for multiple observables. For a single observable the two definitions coincide.\\
Second, the matrices in ref.~\cite{Meyer_2020} are replaced with block matrices, where the individual blocks contain the corresponding quantities for the multiple observables. For example, because molecular dynamics simulations and their results are discrete in time, the (sampled) two-time correlation function (\cref{eq:def_two_time_correlation}) for a single observable can be represented as a matrix, where the row and column specify the two times of the correlation (cf. ref.~\cite{Meyer_2020}). Now, each of these values is replaced by the correlation matrix for multiple observables at the corresponding times. If $\mmat{C}$ denotes the block matrix for the two-time correlation function, we can write it in the following form
\begin{align}
	\mmat{C} &= \begin{pmatrix}
		\mat{C}(0,0) & \cdots&\mat{C}(0,T)\\
		\vdots & \ddots &\vdots\\
		\mat{C}(T,0)&\cdots&\mat{C}(T,T)
	\end{pmatrix}.
\end{align}
Third, the splitting into triangular matrices, as used in ref.~\cite{Meyer_2020}, is done only on the block level. For example, if we want to use the upper-right triangular block matrix of $\mmat{C}$, the blocks on the diagonal ($\mat{C}(0,0)$ etc.) are either kept or discarded (depending on the convention) as a whole.\\

\section{Generalized Langevin dynamics simulations}\label{sec:numerical_method}
The aim of a coarse-graining procedure is to obtain a model, which allows to generate the dynamics of an observable by directly solving the corresponding Langevin equation rather than simulating the underlying microscopic system.
Hence the typical tasks when coarse-graining, are first to determine the drift and memory kernel based on a given set of data from simulations or experiments, and then to generate a random noise, which mimics the fluctuating force.  \\
In this section, we analyze how to simulate generalized Langevin dynamics based on a given data set such that the simulations are statistically equivalent to the original data up to a given order on the level of moments. The main results are necessary and sufficient criteria, which we derive in detail in sec.~\ref{sec:second_order} and sec.~\ref{sec:general_procedure}. In short, we prove that the simulations obey the same dynamics up to a given order, if and only if the direct sum of the fluctuating forces and initial values $(\vec{\eta}^\top_{tt_0},\vec{A}^\top_{t_0})^\top$ are accurately reproduced up to the same order. Note that in general we have to respect the correlations between the initial values and fluctuating forces as demonstrated in sec.~(\ref{ssec:exampleSingleObservable}). The confident reader may skip \cref{sec:second_order,sec:general_procedure} and continue with sec.~(\ref{sec:simplified_procedure}), where we show how to draw the initial values and fluctuating forces independently within the 'second order' approximation.\\
Prior to the derivation, we give some preliminary thoughts and definitions.
Once the drift $\mat{\omega}(t)$ and memory kernel $\mat{K}(t,s)$ are evaluated from a given data-set, we obtain the fluctuating forces $\vec{\eta}_{tt_0}(\Gamma_{t_0})$ for each trajectory $\vec{A}_t(\Gamma_{t_0})$ with initial value $\vec{A}(\Gamma_{t_0})$ from the nsGLE, eq.~(\ref{eq:gle2}). Vice versa, by integrating the nsGLE with the given fluctuating forces and initial value, we recover the trajectory $\vec{A}_t(\Gamma_{t_0})$. Hence, for a given drift $\mat{\omega}(t)$ and kernel $\mat{K}(t,s)$, the phase space density $\rho({t_0},.)$ defines a function of time $t$ and initial value $\Gamma_{t_0}$ via,  
\begin{equation}
	\vec{X}:\, (t,\Gamma_{t_0})\mapsto \vec{X}_t(\Gamma_{t_0})=(\vec{\eta}^\top_{tt_0}(\Gamma_{t_0}),\vec{A}^\top_{t_0}(\Gamma_{t_0}))^\top \, , \label{eq:stochastic_process}
\end{equation}
 which uniquely determines the initial values and time-evolution of the trajectory $\vec{A}_t(\Gamma_{t_0})$. 
In an abstract mathematical sense, $\vec{X}$ is a \textit{continuous-time stochastic process}, where the sample space is the phase space \cite{lamperti1977,bremaud2014}. As a function of solely the sample space, one may consider $\vec{X}$ to be a \textit{random function} that maps some random initial value $\Gamma_{t_0}$ onto a function of time.

 We seek to simulate further trajectories that have dynamics "similar to the dynamics of $\vec{X}$". 
 To this end, we construct a stochastic process $\vec{X}'$ in order to replace eq.~(\ref{eq:stochastic_process}), which approximates the true dynamics on the level of a given set of moments. The first step is to classify all stochastic processes of the form
\begin{equation}
	\vec{X}':\, (t,\Lambda)\mapsto \vec{X}'_t(\Lambda)=(\vec{\eta}'^\top_{tt_0}(\Lambda),\vec{A}'^\top_{t_0}(\Lambda))^\top \, ,
	\label{eq:stochastic_process2}
\end{equation}
that will preserve the mean and auto-correlation function of the trajectories, where $\Lambda$ is the outcome of some sample space specifying the fluctuating forces $\vec{\eta}'_{tt_0}(\Lambda)$ as well as the initial value $\vec{A}'_{t_0}(\Lambda)$. The resulting trajectory obtained by numerical integration is denoted by $\vec{A}'_t(\Lambda)$.

\subsection{Reproducing of the First and Second Moment of the Observables}\label{sec:second_order}

Let us begin with the mean value $\langle \vec{A}_t \rangle$, where $\langle . \rangle$ denotes the expected value of some random variable. For given functions $\mat{\omega}(t)$ and $\mat{K}(t,s)$, the mean value $\langle \vec{A}_t\rangle$ solves the following system of Volterra integro-differential equations
\begin{equation*}
\left. \begin{cases}
\frac{d}{dt} \langle \vec{A}_t \rangle = \mat{\omega}(t)\langle \vec{A}_t\rangle + \int^t_{t_0} ds \, \mat{K}(t,s)\langle \vec{A}_s \rangle + \langle \vec{\eta}_{tt_0} \rangle  \\
\text{initial value: }\, \langle \vec{A}_{t_0} \rangle 
\end{cases}  \right\} \, .
\end{equation*}
For each initial value $\langle \vec{A}_{t_0}\rangle$, these kind of equations possess a unique solution for any continuous functions $\langle \vec{\eta}_{tt_0}\rangle$, $\mat{\omega}(t)$ and $\mat{K}(t,s)$, see ref.~\cite{burton} for details.

If we intend to use \cref{eq:gle2} to generate new trajectories, we use the original drift matrix $\mat{\omega}(t)$ and memory kernel matrix $\mat{K}(t,\tau)$ and  randomly draw the initial values of the observables $\vec{A}'_{t_0}$ and the values of the fluctuating force during the process $\vec{\eta}'_{tt_0}$. From the considerations above we know that the mean values of the generated trajectories equal the original ones, $\langle\vec{A}'_t\rangle\equiv\langle\vec{A}_t\rangle$, if and only if $\langle\vec{A}'_{t_0}\rangle = \langle\vec{A}_{t_0}\rangle$ and $\langle\vec{\eta}'_{tt_0}\rangle\equiv\langle\vec{\eta}_{tt_0}\rangle$. In shorthand notation, these two conditions can also be written as  $\langle\vec{X}'_t\rangle\equiv\langle\vec{X}_t\rangle$.

\begin{widetext}
Now let us continue in the same fashion with the auto-correlation function. Inserting the equations of motion (\cref{eq:gle2}) into the time derivative of the auto-correlation function yields again a system of Volterra differential equations:
\begin{align}
&\left. \begin{cases}
\frac{d}{dt} \langle \vec{A}_t  \vec{A}_s^\top \rangle = \mat{\omega}(t)\cdot \langle \vec{A}_t  \vec{A}_s^\top \rangle + \int^t_{t_0} dr \, \mat{K}(t,r)\cdot\langle \vec{A}_r  \vec{A}_s^\top \rangle + \langle \vec{\eta}_{tt_0}  \vec{A}_s^\top \rangle\\
\text{initial value: }\, \langle \vec{A}_{t_0}   \vec{A}_s^\top \rangle
\end{cases}  \right\} \label{eq:ac1}
\end{align}
Here, the dynamics of $\langle \vec{A}_t  \vec{A}_s^\top \rangle$ for arbitrary but fixed $s$ are completely determined by the correlation function $\langle \vec{\eta}_{tt_0}  \vec{A}_s^\top \rangle$ and the initial value $\langle \vec{A}_{t_0}   \vec{A}_s^\top \rangle$. We want to understand under which conditions of our simulation a) the correlation function $\langle \vec{\eta}_{tt_0}  \vec{A}_s^\top \rangle$ and b) the initial value $\langle \vec{A}_{t_0}   \vec{A}_s^\top \rangle$ (for arbitrary $s$) show the same behavior as in the original simulations of the microscopic system:
\begin{itemize}
    \item[a)] We start with the correlation function $\langle \vec{\eta}_{tt_0}  \vec{A}_s^\top \rangle$ and derive an equation of motion for it by calculating the derivative with respect to $s$ and using the nsGLE \cref{eq:gle2} again. The result reads:
\begin{align}
&\left. \begin{cases}
\frac{d}{ds}\langle \vec{\eta}_{tt_0}  \vec{A}_s^\top \rangle = \langle \vec{\eta}_{tt_0}  \vec{A}_s^\top \rangle \cdot \mat{\omega}(s)^\top + \int^s_{t_0} dr\,  \langle \vec{\eta}_{tt_0} \vec{A}_r^\top \rangle \cdot \mat{K}(t,r)^\top + \langle \vec{\eta}_{tt_0}  \vec{\eta}_{st_0}^\top\rangle\\
\text{initial value: }\, \langle \vec{\eta}_{tt_0}  \vec{A}_{t_0}^\top \rangle
\end{cases}  \right\} \label{eq:ac2}
\end{align}
Hence, the dynamics of $\langle \vec{\eta}_{tt_0}  \vec{A}_s^\top \rangle$ for fixed $t$ is, again, completely determined by the initial value $\langle \vec{\eta}_{tt_0}  \vec{A}_{t_0}^\top \rangle$ and the two-time correlation of the fluctuating force $\langle \vec{\eta}_{tt_0}  \vec{\eta}_{st_0}^\top\rangle$.

\item[b)] Next, we ask under which conditions the initial value $\langle \vec{A}_t   \vec{A}_s^\top \rangle\rvert_{t={t_0}}$ (for arbitrary $s$) has the same form as in the original microscopic simulations. Again, we use a time derivative with respect to $s$ and the nsGLE  and obtain:
\begin{align}
&\left. \begin{cases}
\frac{d}{ds}\langle \vec{A}_{t_0}  \vec{A}_s^\top \rangle = \langle \vec{A}_{t_0}  \vec{A}_s^\top \rangle\cdot \mat{\omega}(s)^\top + \int^s_{t_0} dr\, \langle \vec{A}_{t_0} \vec{A}_r^\top \rangle\cdot \mat{K}(t,r)^\top + \langle \vec{A}_{t_0}  \vec{\eta}_{st_0}^\top\rangle\\
\text{initial value: }\, \langle \vec{A}_{t_0}  \vec{A}_{t_0}^\top \rangle
\end{cases}  \right\} \label{eq:ac3}\, . 
\end{align}
Here we can see that $\langle \vec{A}_{t_0}  \vec{A}_s^\top \rangle$ is completely determined by the initial value $\langle \vec{A}_{t_0}  \vec{A}_{t_0}^\top \rangle$ and the correlation function $\langle \vec{A}_{t_0}  \vec{\eta}_{st_0}^\top\rangle$.
\end{itemize}
\end{widetext}

Let us summarize the results above in the following way:\\
If the dynamics of our observables are described by an equation of the structure of the nsGLE (\cref{eq:gle2}), the two-time correlation functions are uniquely determined if $\langle \vec{A}_{t_0}  \vec{A}_{t_0}^\top\rangle$, $\langle \vec{A}_{t_0}  \vec{\eta}_{st_0}^\top \rangle$, and $\langle \vec{\eta}_{tt_0} \vec{\eta}_{st_0}^\top\rangle$ are given for all times $t$ and $s$. In compact form, these three correlation functions can be expressed as $\langle \vec{X}_t\vec{X}_s^\top\rangle$. Here, $\langle \vec{A}_{t_0}  \vec{A}_{t_0}^\top\rangle$ and $\langle \vec{A}_{t_0}  \vec{\eta}_{st_0}^\top \rangle$ determine $\langle \vec{A}_{t_0}  \vec{A}_s^\top\rangle$ via \cref{eq:ac3} and similarly $\langle \vec{A}_{t_0}  \vec{\eta}_{st_0}^\top \rangle$ and $\langle \vec{\eta}_{tt_0} \vec{\eta}_{st_0}^\top\rangle$ determine $\langle \vec{\eta}_{tt_0} \vec{A}_s^\top\rangle$ via \cref{eq:ac2}. With all these quantities uniquely determined, also the two-time correlation function $\mat{C}(t,s)=\langle \vec{A}_t \vec{A}_s^\top\rangle$ is uniquely determined through \cref{eq:ac1}. The special case where $\langle \vec{A}_{t_0}  \vec{\eta}_{st_0}^\top\rangle\equiv 0$, as is the case for the nsGLE with  \cref{eq:mean_f_A}, is discussed in \cref{sec:simplified_procedure}.

We conclude that new trajectories generated with the nsGLE (\cref{eq:gle2}) also reproduce the exact two-time auto-correlation functions, $\langle\vec{A}'_t\vec{A}'_s\rangle\equiv\langle\vec{A}_t\vec{A}_s\rangle$, if and only if $\langle\vec{X}'_t\vec{X}'_s\rangle\equiv\langle\vec{X}_t\vec{X}_s\rangle$. These results may be used to simulate trajectories that obey the same statistics up to second order as we will illustrate in \cref{sec:simplified_procedure}.

\subsection{Reproducing Higher Moments of the Observables}\label{sec:general_procedure}
In section \ref{sec:second_order}, we have seen that the simulated trajectories obey the same statistics up to second order, if and only if the stochastic process $\vec{X}'$ from \cref{eq:stochastic_process2} preserves the first and second moment of the true process $\vec{X}$ from \cref{eq:stochastic_process}. In fact, the same line of argumentation applies to arbitrary orders. To this end we define the $m$-th moment by
\begin{align}
	\mat{C}(t_1,\dots,t_m)&:=\langle \vec{A}_{t_1}\otimes\dots\otimes \vec{A}_{t_m}\rangle \, .
\end{align}

\begin{widetext}
Again, we use the nsGLE (\cref{eq:gle2}) to express the $m$-th moment by the unique solution of an initial value problem. Then, we recursively repeat the procedure until all terms are determined by the $m$-th moment of the random process $\vec{X}$. We obtain
\begin{align}
&\left. \begin{cases}
\frac{d}{dt_1} \mat{C}(t_1,\dots,t_m) = \langle \vec{\eta}_{t_1t_0} \otimes \vec{A}_{t_2} \otimes \cdots \otimes \vec{A}_{t_m} \rangle + (\mat{\omega}(t_1) [\otimes\mat{\mathbf{1}}]^{m-1} )\cdot \mat{C}(t_1,\dots,t_m) + \\
\qquad + \int^t_{t_0} ds \, (\mat{K}(t,s)[\otimes\mat{\mathbf{1}}]^{m-1})\cdot \mat{C}(s,t_2,\dots,t_m)  \\
\text{initial value: }\, \langle \vec{A}_{{t_0}}\otimes \vec{A}_{t_2}\otimes\dots\otimes \vec{A}_{t_m}\rangle
\end{cases} \right\} \, . \label{eq:M_th_moment_1}
\end{align}
As before, we can show that the solution to this integro-differential equation is uniquely determined by the following set of integro-diffential equations:
\begin{align}
\left.\begin{cases}
\frac{\dd}{\dd t_2} \langle \vec{X}_{t_1}\otimes \vec{A}_{t_{2}}\otimes \cdots \otimes \vec{A}_{t_m} \rangle = \langle \vec{X}_{t_1}\otimes\vec{\eta}_{t_{2}t_0}\otimes \vec{A}_{t_{k+1}}\otimes \cdots \otimes \vec{A}_{t_m} \rangle + \\
\qquad+ (\mat{\mathbf{1}}\otimes\mat{\omega}(t_2) [\otimes\mat{\mathbf{1}}]^{m-2}) \cdot \langle \vec{X}_{t_1}\otimes\vec{A}_{t_{2}}\otimes \cdots \otimes \vec{A}_{t_m} \rangle + \\
\qquad+  \int^t_{t_0} ds \, (\mat{\mathbf{1}}\otimes \mat{K}(t_k,s) [\otimes\mat{\mathbf{1}}]^{m-2} )\cdot \langle \vec{X}_{t_1}\otimes \vec{A}_{s}\otimes \vec{A}_{t_{3}}\otimes \cdots \otimes \vec{A}_{t_m} \rangle  \\
\text{initial value: }\, \langle \vec{X}_{t_1}\otimes \vec{A}_{{t_0}}\otimes \vec{A}_{t_{2}}\otimes \cdots \otimes \vec{A}_{t_m} \rangle
\end{cases} \right\} \, , \label{eq:M_th_moment_2} 
\end{align}
This notation allows us to write the generalization of both \cref{eq:ac2,eq:ac3} in the single expression of \cref{eq:M_th_moment_2}. If $m=2$, \cref{eq:ac2,eq:ac3} are actually equivalent to \cref{eq:M_th_moment_2}.\\
Now, we can perform the same procedure iteratively and after $k$ iterations we obtain:
\begin{align}
\left.\begin{cases}
\frac{d}{dt_k} \langle \vec{X}_{t_1}\otimes\cdots\otimes \vec{X}_{t_{k-1}}\otimes \vec{A}_{t_{k}}\otimes \cdots \otimes \vec{A}_{t_m} \rangle = \langle \vec{X}_{t_1}\otimes\cdots\otimes \vec{X}_{t_{k-1}}\otimes \vec{\eta}_{t_{k}t_0}\otimes \vec{A}_{t_{k+1}}\otimes \cdots \otimes \vec{A}_{t_m} \rangle + \\
\qquad+ ([\mat{\mathbf{1}}\otimes]^{k-1} \mat{\omega}(t_k) [\otimes\mat{\mathbf{1}}]^{m-k} )\cdot \langle \vec{X}_{t_1}\otimes\cdots\otimes \vec{X}_{t_{k-1}}\otimes \vec{A}_{t_{k}}\otimes \cdots \otimes \vec{A}_{t_m} \rangle + \\
\qquad+  \int^t_{t_0} ds \, ([\mat{\mathbf{1}}\otimes]^{k-1} \mat{K}(t_k,s) [\otimes\mat{\mathbf{1}}]^{m-k} )\cdot \langle \vec{X}_{t_1}\otimes\cdots\otimes \vec{X}_{t_{k-1}}\otimes \vec{A}_{s}\otimes \vec{A}_{t_{k+1}}\otimes \cdots \otimes \vec{A}_{t_m} \rangle  \\
\text{initial value: }\, \langle \vec{X}_{t_1}\otimes\cdots\otimes \vec{X}_{t_{k-1}}\otimes \vec{A}_{{t_0}}\otimes \vec{A}_{t_{k+1}}\otimes \cdots \otimes \vec{A}_{t_m} \rangle
\end{cases} \right\} \, , \label{eq:M_th_moment_k} 
\end{align}
\end{widetext}

In principle, we can now recursively solve \cref{eq:M_th_moment_k} for $k=m,m-1,\dots,1$ in order to obtain the unique solution $\mat{C}(t_1,\cdots,t_m)$, if we are given the $m$-th moment of the original dynamics $\langle \bigotimes^m_{k=1} \vec{X}_{t_k}\rangle$ for all times. Hence, if and only if the $m$-th moments of $\vec{X}'_t$ in the stochastic process equal the original $m$-th moments of $\vec{X}_t$, the $m$-th moments of the observables are reproduced. More explicitly,
\begin{equation*}
	\left\langle \bigotimes^m_{k=1}\vec{A}'_{t_k}  \right\rangle \equiv  \left\langle \bigotimes^m_{k=1}\vec{A}_{t_k}\right\rangle \! \Leftrightarrow \! \left\langle \bigotimes^m_{k=1}\vec{X}'_{t_k}\right\rangle \equiv  \left\langle \bigotimes^m_{k=1}\vec{X}_{t_k}\right\rangle  ,
\end{equation*}
for any $m\in\mathbb{N}_+$. Therefore, if we choose $\vec{X}'$ such that $\left\langle \bigotimes^m_{k=1}\vec{X}'_{t_k}\right\rangle \equiv  \left\langle \bigotimes^m_{k=1}\vec{X}_{t_k}\right\rangle$ for all times and $m=1,\dots,M$, we obtain simulations that obey the same statistics up to $M$-th order by numerical integration of the nsGLE (\ref{eq:gle2}) for any given functions $\mat{\omega}(t)$ and $\mat{K}(t,s)$.

\subsection{Simplified procedure}\label{sec:simplified_procedure}

While the general procedure is independent from the construction of the projection operator formalism, we can now exploit \cref{eq:mean_f_A} in order to bypass statistical dependencies between the fluctuating forces and initial conditions within the second order approximation. 

Let us demand $\langle \vec{A}_{t_0}\rangle=0$, i.e. we subtract the average initial value $\vec{A}_t\to \vec{A}_t-\langle \vec{A}_{t_0}\rangle$, which is easily reversed after the numerical integration is carried out. Further, we assume that $\vec{\eta}'_{tt_0}$ and $\vec{A}'_{t_0}$ are uncorrelated random variables. According to \cref{eq:mean_f_A}, we have $\langle \vec{\eta}_{st_0}  \vec{A}_{t_0}^\top\rangle\equiv0$. Hence, if $\langle \vec{A}'_{t_0}\rangle =0$, we automatically satisfy 
\begin{equation*}
\langle \vec{\eta}'_{st_0} \vec{A}'^\top_{t_0}\rangle = \langle \vec{\eta}'_{st_0}\rangle  \langle \vec{A}'^\top_{t_0}\rangle = 0 = \langle \vec{\eta}_{st_0} \vec{A}^\top_{t_0}\rangle \, ,
\end{equation*}
for all times $s$. From this, we conclude the following statement. Let $\langle \vec{A}_{t_0}\rangle=0$ and let $\vec{\eta}'_{tt_0}$, $\vec{A}'_{t_0}$ be uncorrelated random variables. Then, $\langle \vec{A}'_t\rangle\equiv \langle \vec{A}_t\rangle$ and $\langle \vec{A}'_t \vec{A}'^\top_s\rangle\equiv\langle \vec{A}_t \vec{A}_s^\top\rangle$, if and only if $\langle \vec{\eta}'_{tt_0}\rangle\equiv \langle \vec{\eta}_{tt_0}\rangle $, $\langle \vec{A}'_{t_0}\rangle=0$ (or $\langle \vec{\eta}'_{tt_0}\rangle\equiv0$), $\langle \vec{\eta}'_{tt_0} \vec{\eta}'^\top_{st_0}\rangle\equiv \langle \vec{\eta}_{tt_0} \vec{\eta}^\top_{st_0}\rangle $, and $\langle \vec{A}'_{t_0}  \vec{A}'^\top_{t_0} \rangle=\langle \vec{A}_{t_0}  \vec{A}^\top_{t_0} \rangle$.  

This suggests the following procedure. Assume we are given a set of trajectories $\vec{B}_t$. First, we compute $\vec{A}_t := \vec{B}_t-\langle \vec{B}_{t_0}\rangle$ for all trajectories. Afterwards, we determine $\mat{\omega}(t)$, $\mat{K}(t,s)$, $\langle \vec{A}_{t_0}  \vec{A}^\top_{t_0} \rangle$, $\langle \vec{\eta}_{tt_0}\rangle$ and $\langle \vec{\eta}_{tt_0} \vec{\eta}_{st_0}^\top \rangle$. For each simulation $\vec{A}'_t$, we draw the initial value $\vec{A}'_{t_0}$ from any distribution with mean zero and second moment $\langle \vec{A}_{t_0}  \vec{A}^\top_{t_0} \rangle$. Then, we integrate the equations of motion, where the fluctuating forces $\vec{\eta}'_{tt_0}$ are drawn from any distribution with mean $\langle \vec{\eta}_{tt_0}\rangle$ and auto-correlation function $\langle \vec{\eta}_{tt_0} \vec{\eta}_{st_0}^\top \rangle$. In the end, the desired simulations are given by $\vec{B}'_t = \vec{A}'_t+\langle \vec{B}_{t_0}\rangle$. Numeric evidence for the necessity of subtracting the mean in the beginning and adding it again in the end is presented later in \cref{ssec:exampleSingleObservable} (see \cref{fig:short_simulations_no_offset_comparison,fig:short_simulations_with_offset_comparison}). There, we also discuss a case where such a procedure is not necessary (see \cref{fig:long_simulations_comparison}).

\section{Numerical Results for Exemplary System}
In this section we will discuss algorithmic/numerical details in \cref{ssec:generatingTrajectories}, introduce a model system in \cref{ssec:exemplarySystem}, present some results for a single observable to illustrate the importance of the addition/subtraction of the mean in the beginning/end in \cref{ssec:exampleSingleObservable}, show the results of a simulation with two observables in \cref{ssec:exampleTwoObservables}, and show an example of how this method can be used to generate new trajectories for system parameters where no original data is available in \cref{ssec:interpolationSchemes}.

\subsection{Generating Trajectories with the Non-Stationary Generalized Langevin Equation}\label{ssec:generatingTrajectories}
In order to simulate trajectories with the nsGLE (\cref{eq:gle2}), we need to generate fluctuating forces with a certain mean $\langle\vec{\eta}_{tt_0}\rangle$ and auto-correlation function $\langle\vec{\eta}_{tt_0}\vec{\eta}_{st_0}\rangle$.
As the simulations are discrete in time, we apply the numerical method to the fluctuating forces $\vec{\eta}_{tt_0}(\Gamma_{t_0})$, where $t=1,\dots,N$ enumerates $N$ instants in time on an equidistant grid. As we care only about the first two moments in this section, a multi-dimensional Gaussian distribution is the natural choice, which is also easy to handle from a computational perspective. For non-stationary processes, we have to allow for arbitrary mean values and covariances, thus, the fluctuating forces are drawn from a $(nN)$-dimensional Gaussian distribution
\begin{align*}
	\vec{\eta}&:=(\vec{\eta}^\top_{1t_0},\dots,\vec{\eta}^\top_{Nt_0})^\top(\Gamma_{t_0}) \in \mathcal{N}(\vec{\mu},\mat{\Sigma}) \, , \\
	\vec{\mu} &= \langle \vec{\eta} \rangle  \, , \\
	\mat{\Sigma} &= \langle (\vec{\eta}-\vec{\mu})(\vec{\eta}-\vec{\mu})^\top\rangle \, .
\end{align*}
In order to obtain the correct distribution of initial values of the observables, the initial values $\vec{A}_{t_0}$ for the $i$-th simulation are simply taken from the ($i\% N_{\text{data}}$)-th trajectory from the given set of $N_{\text{data}}$ original trajectories. The numerical integration is carried out by the classical Runge-Kutta method with $dt=2$, where we use the Simpson rule to compute the integral from the nsGLE. Odd times are restored using the Verlet scheme after each Runge-Kutta step at the cost of one order in accuracy, i.e. the local error is in $O(1/N^4)$, assuming that $\vec{\eta}_{tt_0}$, $\mat{\omega}(t)$ and $\mat{K}(t,s)$ are accurate and sufficiently smooth. 

\subsection{Exemplary System: Dipole Gas}\label{ssec:exemplarySystem}
As a simple example to illustrate the methods derived above, we consider the dynamics of the total dipole moment of a dilute gas of magnetic dipoles (cf. ref.~\cite{Meyer_2021}). The system consists of $N_\text{p}=125$ particles in a three dimensional box with periodic boundary conditions. The individual particles interact via a steric Weeks-Chandler-Anderson interaction with an interaction potential of the form
\begin{align}
	V_{\text{WCA}}(r) &= \begin{cases}
	4\epsilon\left[\left(\frac{\sigma}{r}\right)^{12}-\left(\frac{\sigma}{r}\right)^{6}+1\right] & \text{if }r<2^{1/6}\sigma\\
	0 & \text{else}
	\end{cases}
\end{align}
and an exact magnetic dipole-dipole interaction \cite{Landecker1900} with the forces $F_{ij}$ and torques $G_{ij}$ on particle $j$ due to particle $i$
\begin{align}
	\vec{F}_{ij} &= \frac{3\mu_0}{4\pi r^4}\left(\left(\vec{n}\times\vec{m}_i\right)\times\vec{m}_j+\left(\vec{n}\times\vec{m}_j\right)\times\vec{m}_i\right.\nonumber\\
	&\phantom{=}\left. -2\vec{n}\left(\vec{m}_i\cdot\vec{m}_j\right)+5\vec{n}\left(\vec{n}\times\vec{m}_i\right)\cdot\left(\vec{n}\times\vec{m}_j\right) \right)\\
	\vec{\mathcal{G}}_{ij} &= \frac{\mu_0}{4\pi r^3}\left(3\left(\vec{m}_i\cdot\vec{n}\right)\left(\vec{m}_j\times\vec{n}\right)+\left(\vec{m}_i\times\vec{m}_j\right)\right)\,.
\end{align}
In these equations $\epsilon$ and $\sigma$ are the usual Lennard-Jones parameters, $r$ is the distance between the two particles, $\vec{n}$ is the 3D vector pointing from particle $i$ to particle $j$, $\mu_0$ is the permeability of free space, and $\vec{m}_{k}$ is the 3D dipole moment. All the interactions are evaluated in the nearest image convention. Further, we include an external homogeneous magnetic field $\vec{B}$, which couples to the magnetic dipoles via an additional torque
\begin{align}
    \vec{\mathcal{G}}_j &= \vec{m}_j\times\vec{B}\,.
\end{align}
For the integration of the equations of motion the velocity-Verlet method is used.

\subsection{Relaxation of a Single Observable}\label{ssec:exampleSingleObservable}

\begin{figure}[hb!]
    \centering
    \includegraphics{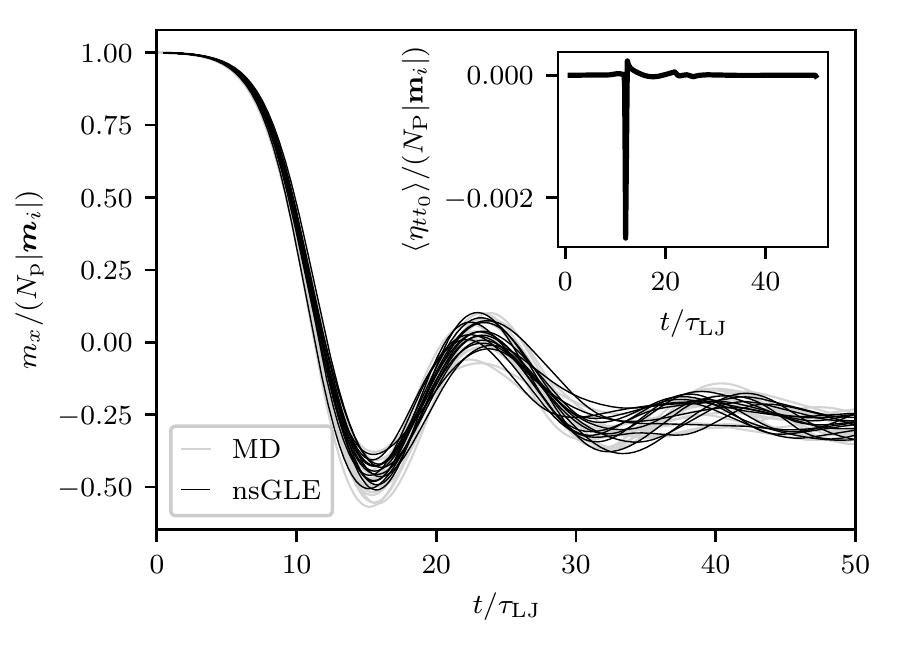}
    \caption{Relaxation of the total magnetic dipole moment along the $x$ direction as a function of time. The gray curves are obtained from MD simulations, the black curves from the nsGLE without the initial shift. The nsGLE simulations start at time $t_0=0.5\tau_\text{LJ}$ when there is almost no spreading of the values of $m_x$. Accordingly, the magnitude of the average of the fluctuating forces, shown in the inset, is comparably small in this case.}
    \label{fig:long_simulations_comparison}
\end{figure}

\begin{figure}[hb!]
    \centering
    \includegraphics{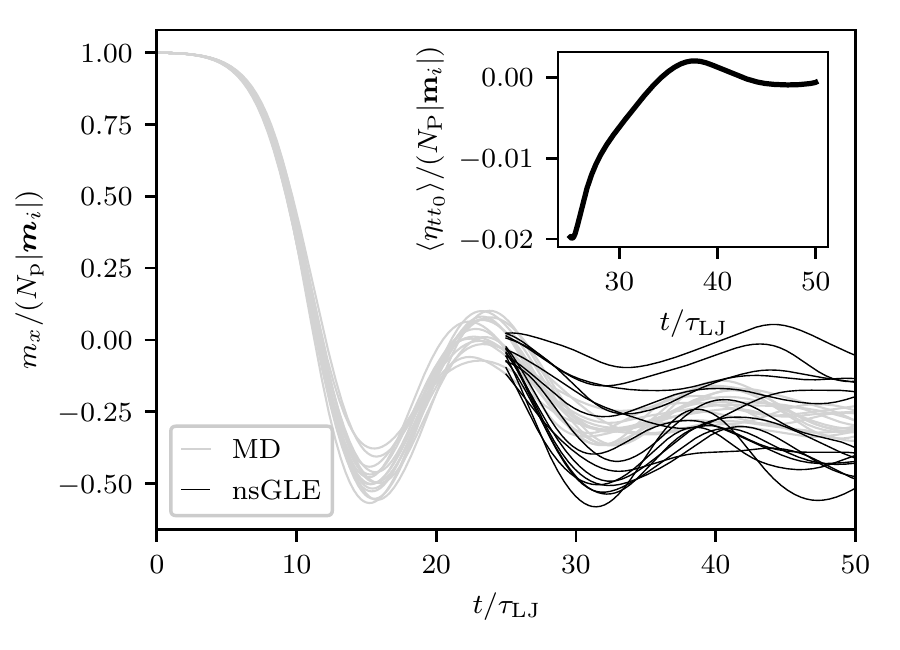}
    \caption{Dynamics of the total magnetic dipole moment along the $x$ direction as a function of time. The gray curves are obtained from MD simulations and the black curves from the nsGLE as described in sec.~\ref{sec:simplified_procedure} and \ref{ssec:generatingTrajectories}, but without the initial shift. Hence, statistical dependencies between the fluctuating forces and initial values are neglected. The nsGLE simulations start at time $t_0=25\tau_\text{LJ}$ when there is a considerable amount of spreading of the values of $m_x$. The inset shows the average of the fluctuating forces, which are at least one order of magnitude larger than the ones in \cref{fig:long_simulations_comparison}.}
    \label{fig:short_simulations_no_offset_comparison}
\end{figure}

\begin{figure}
    \centering
    \includegraphics{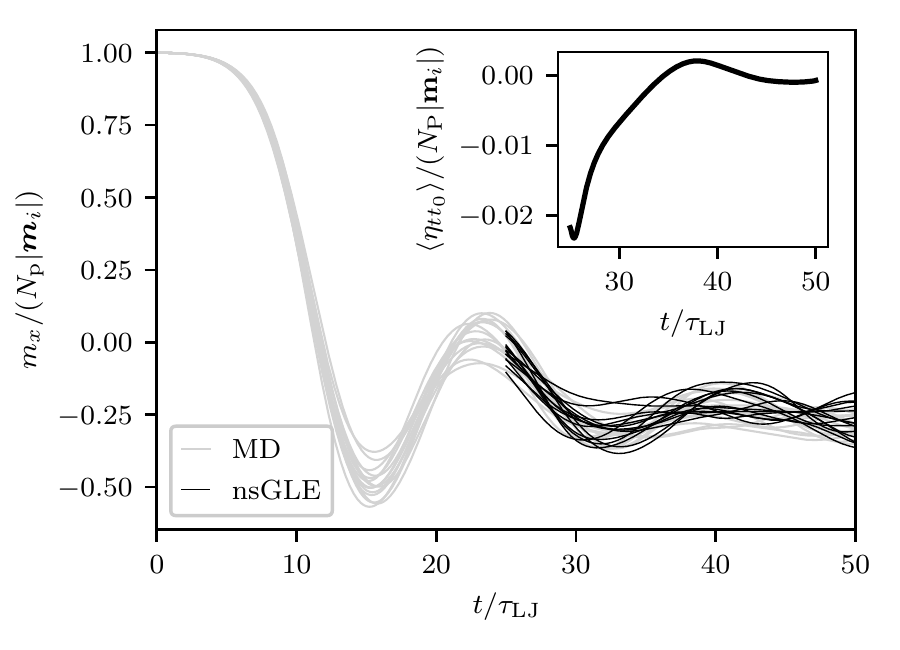}
    \caption{Dynamics of the total magnetic dipole moment along the $x$ direction as a function of time. The gray curves are obtained from MD simulations, the black curves from the nsGLE with an initial shift $\vec{A}_t\mapsto\vec{A}_t-\langle\vec{A}_{t_0}\rangle$. The nsGLE simulations start at time $t_0=25\tau_\text{LJ}$ when there is a considerable amount of spreading of the values of $m_x$. Due to the shift, the trajectories generated by the nsGLE still reproduce the original dynamics well. The inset shows the average of the fluctuating forces, which are at least one order of magnitude larger than the ones in \cref{fig:long_simulations_comparison}.}
    \label{fig:short_simulations_with_offset_comparison}
\end{figure}
First we investigate the relaxation of the dipole moment along one direction. For this purpose we initialize the system with all particles positioned on a simple cubic lattice and all magnetic dipoles aligned along the positive $x$-direction. Further, an external magnetic field pointing along the negative $x$-direction is added. Hence, the dipoles minimize their potential energy with respect to the external field if they rotate $180^\circ$. To get the system out of the unstable equilibrium state, the particles are initialized with some random (angular) momenta, that are small compared to the values they take during the rest of the simulation. \\
All particles have a mass $m$ and a moment of inertia of $I=m\sigma^2/4$. They are placed in a box of size $(6\sigma)^3$ and the strength of the dipole-dipole interaction is set through the relation $\mu_0\vec{m}_i^2=\sigma^3\epsilon/8$. The strength of the external magnetic field is $|\vec{B}|=0.01\epsilon/|\vec{m}_i|$. In the following plots, the time is given in units of $\tau_\text{LJ}=\sqrt{m\sigma^2/\epsilon}$. In total, 1000 simulations of duration $T_\text{max}=50\tau_\text{LJ}$ are run. Here, the coarse-grained observable is the $x$-component of the total dipole moment $\vec{m}=\sum_i\vec{m}_i$. Ten example trajectories are shown as gray lines in \cref{fig:long_simulations_comparison,fig:short_simulations_no_offset_comparison,fig:short_simulations_with_offset_comparison}.\\
In \cref{fig:long_simulations_comparison}, 10 example trajectories generated with the nsGLE are shown in comparison to the original MD trajectories. Here, the observable has not been shifted according to the procedure described in sec.~\ref{sec:simplified_procedure}. Nevertheless, the generated trajectories agree with the original MD ones very well. Recall that the shift in the beginning and end is done to ensure that $\langle \vec{\eta}_{tt_0} \vec{A}^\top_{t_0}\rangle\equiv \mat{0}$ for all $t$. However, if the spread of the observable in the beginning is very small, as is the case in \cref{fig:long_simulations_comparison}, the value of the observable is almost the same $\vec{A}_{t_0}\approx\left\langle\vec{A}_{t_0}\right\rangle$ and we find that the mean of the fluctuating force has to vanish:
\begin{align*}
    \mat{0} &\equiv\left\langle \vec{A}_{t_0}\vec{\eta}^\top_{tt_0}\right\rangle\\
    &\approx \left\langle\vec{A}_{t_0}\right\rangle \left\langle \vec{\eta}^\top_{tt_0}\right\rangle\\
    \Rightarrow \quad \forall_t \,:\quad  \left\langle\vec{\eta}_{tt_0}\right\rangle &\approx \vec{0}\label{eq:ff_is_zero}  \, .
\end{align*}
Hence, in this case, the necessary condition $\langle \vec{\eta}_{tt_0} \vec{A}^\top_{t_0}\rangle \equiv 0$ is approximately satisfied for the generated trajectories as well, even without shifting the trajectories. Indeed, in \cref{fig:long_simulations_comparison}, the average of the fluctuating force term is very small for almost all $t$, hence, the trajectories generated with the nsGLE look very good even without the shifts. 

In \cref{fig:short_simulations_no_offset_comparison} we show a counter example. There, the nsGLE simulations are started at time $t=25\tau_\text{LJ}$ when there is already a considerable spreading of the trajectories. Accordingly, the trajectories generated with the nsGLE without the shifts  do not agree with the original ones at all. However, if one adds the shifts, as done in \cref{fig:short_simulations_with_offset_comparison}, the trajectories generated with the nsGLE and the original ones show the same behavior again.

\subsection{Dynamics of a Set of Two Observables}\label{ssec:exampleTwoObservables}
\begin{figure}
    \centering
    \includegraphics{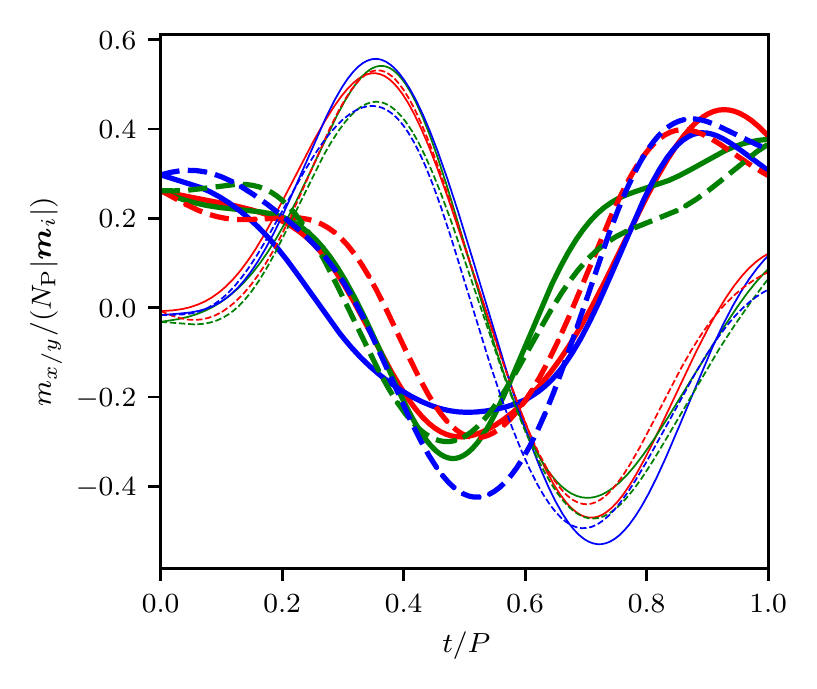}
    \caption{Exemplary trajectories of the total dipole moment in $x$-direction (bold lines) and $y$-direction (thin lines) for a period of $P=35\tau_\text{LJ}$. The continuous lines are obtained from the MD simulations, the dashed ones are obtained using the nsGLE. The color indicates the set of initial values of the observables used for the integration.}
    \label{fig:exemplaryTrajectories}
\end{figure}
\begin{figure}
    \centering
    \includegraphics{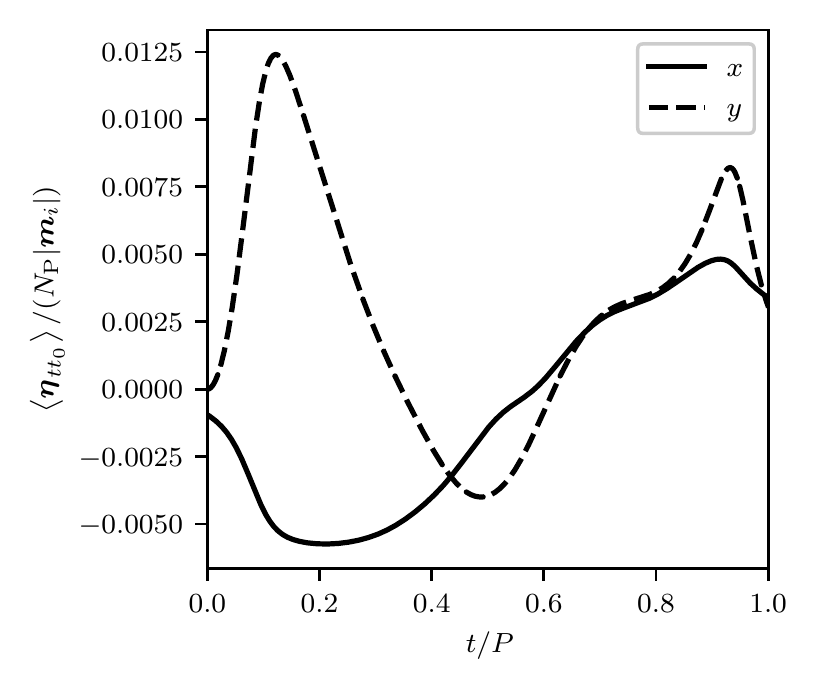}
    \caption{Components of the average of the fluctuating force for a period of $P=35\tau_\text{LJ}$.}
    \label{fig:ff_average70}
\end{figure}

In this section, we use the formalism derived above to describe the dynamics of two correlated observables, namely the total dipole moment along the $x$- and $y$-direction. To obtain interesting dynamics for both observables, we apply the following modifications:\\
We place the particles in a larger simulation box of size $(50\sigma)^3$ and initialize all trajectories with a constant magnetic field $\vec{B}=0.01\epsilon\vec{e}_x/|\vec{m}_i|$. After a short simulation time of $100\tau_\text{LJ}$ with a Nos\'e-Hoover-chain thermostat (target temperature of $T=1\epsilon$, assuming the Boltzmann constant $k_\text{B}=1$), the thermostat is turned off and the external field (with unaltered strength) starts rotating in the $xy$ plane
\begin{align*}
    \vec{B}(t) &= B\left(\cos\left(2\pi\frac{t-t_0}{P}\right)\vec{e}_x+\sin\left(2\pi\frac{t-t_0}{P}\right)\vec{e}_y\right),
\end{align*}
where $P$ is the period of the rotation. At the time the rotation starts, we start sampling the total magnetic dipole moment $\vec{m}$. The time-step for the simulation is $\delta t=0.005\tau_{LJ}$. For every period $P\in\{25\tau_\text{LJ},30\tau_\text{LJ},32.5\tau_\text{LJ},35\tau_\text{LJ}\}$ we sample 1000 trajectories.

In the following, the observables of interest are the total dipole moment in $x$- and $y$-direction $\vec{A}=(m_x,m_y)^\top$. Some exemplary trajectories from the molecular dynamics (MD) simulations as well as some trajectories generated with the nsGLE can be seen in \cref{fig:exemplaryTrajectories}.

\begin{figure}
    \centering
    \includegraphics{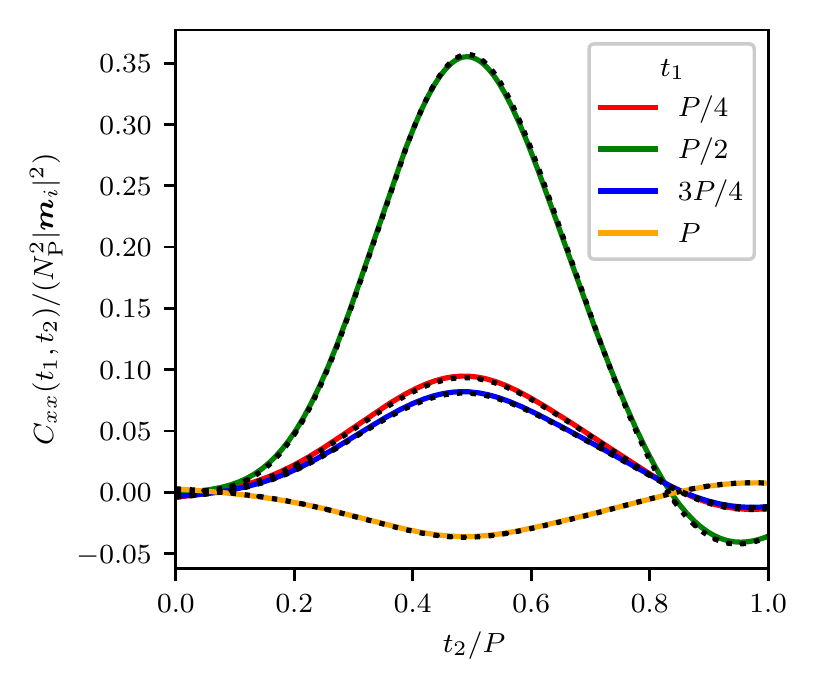}
    \caption{Slices of the two-time correlation function from the original MD data (continuous colored lines) and the nsGLE data (dotted black lines). Here, the $xx$-component (corresponding to the auto-correlation of $m_x$) is shown.}
    \label{fig:CorrelationComparison}
\end{figure}

The quantities shown in \cref{fig:ff_average70,fig:CorrelationComparison} are calculated from data that is shifted in the beginning and end.
In \cref{fig:ff_average70} the average of the fluctuating force is shown, which is non-zero for almost all times. 
In \cref{fig:CorrelationComparison} a comparison of the two-time correlation function calculated from the original MD data and from the nsGLE data can be seen. As expected, they agree perfectly.

\subsection{Interpolation Schemes}\label{ssec:interpolationSchemes}
If the method presented here only allowed to reproduce dynamics, which had been obtained by means of a simulation of the original, microscopic system, it would not be of much use. It turns into a proper coarse-graining method, once the drift and memory kernel can be extrapolated (or interpolated) to conditions other than those originally simulated. Here we show how to obtain such an interpolation.

In many cases, the memory kernels of the nsGLE show a simple exponential or power law behavior. In these cases, one might try to parametrize the shape of the memory kernel in terms of system parameters (e.g. the period of an external field) and use the methods introduced in this article in order to efficiently generate data for new parameter sets. The memory kernels for our example system do not have such a simple shape. However, using some physical intuition we can interpolate from two existing data sets (here, periods of the external field of $P=30\tau_\text{LJ}$ and $P=35\tau_\text{LJ}$) to generate trajectories for new system parameters (here, a period of the external field of $P32.5\tau_\text{LJ}$).

For the interpolation we use the intuition, that the dynamics of our observables are driven by the phase (or direction) of the external magnetic field. Thus, we interpolate between values corresponding to the same phase of the external field. Because the external field rotates with a constant angular frequency, the external phase is proportional to $t/P$ and we use the following linear interpolation for the memory kernel 
\begin{align*}
    \mat{K}^{(2)}(t,s) &:= (1-\lambda) \mat{K}^{(1)}(t P_1/P_2,s P_1/P_2)+ \\
    &\quad +\lambda\mat{K}^{(3)}(t P_3/P_2,s P_3/P_2) \, ,
\end{align*}
in order to interpolate between the periods $P_1$ and $P_3$, where $P_2=(1-\lambda)P_1+\lambda P_3$. The interpolation for the coefficients of the fluctuating force distribution is carried out analogously.

\begin{figure*}
    \centering
    \includegraphics[]{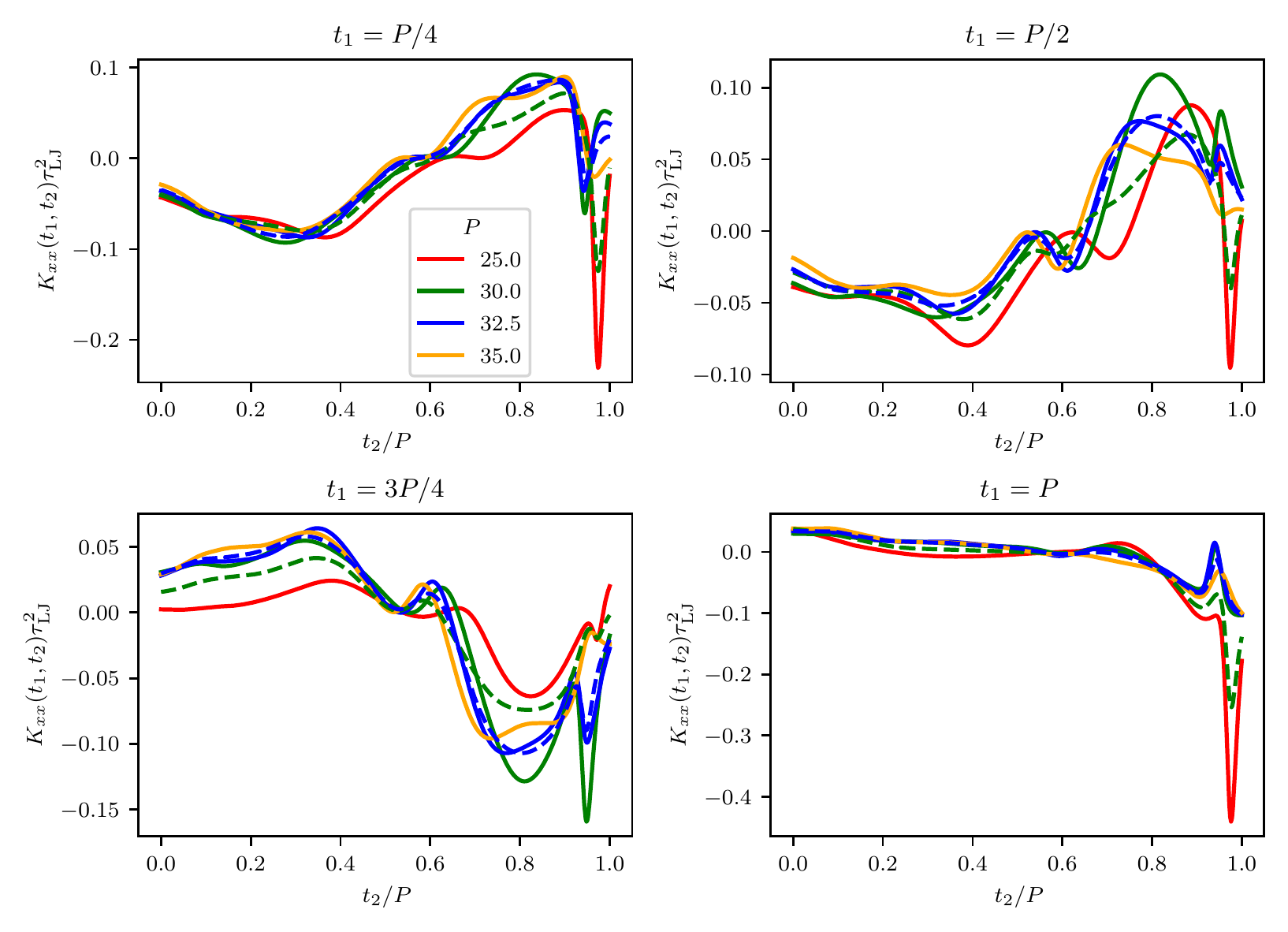}
    \caption{Exemplary slices of the $xx$ component of memory kernels obtained from MD simulations (solid lines) and obtained from interpolation (dashed lines). The different colors correspond to different periods of the external field. The dashed blue line (for $P=32.5\tau_\text{LJ}$) is obtained by taking the mean of the solid green ($P=30\tau_\text{LJ}$) and the solid orange ($P=35\tau_\text{LJ}$) line. The dashed green line (for $P=30\tau_\text{LJ}$) is obtained by taking the mean of the solid red ($P=25\tau_\text{LJ}$) and the solid orange ($P=35\tau_\text{LJ}$) line.}
    \label{fig:kernelInterpolation}
\end{figure*}

In \cref{fig:kernelInterpolation} some exemplary slices of the $xx$ component of memory kernels for different periods are shown both for the original MD data and the interpolation. As can be seen, the mean of the memory kernel for $P=30\tau_\text{LJ}$ and $P=35\tau_\text{LJ}$ is close to the real memory kernel for $P=32.5\tau_\text{LJ}$. However, the interpolation from $P=25\tau_\text{LJ}$ and $P=35\tau_\text{LJ}$ to $P=30\tau_\text{LJ}$ is quite far off from the actual memory kernel for $P=30\tau_\text{LJ}$ and, hence, this is definitely beyond the regime where a simple linear interpolation can be done. 

\begin{figure}
    \centering
    \includegraphics{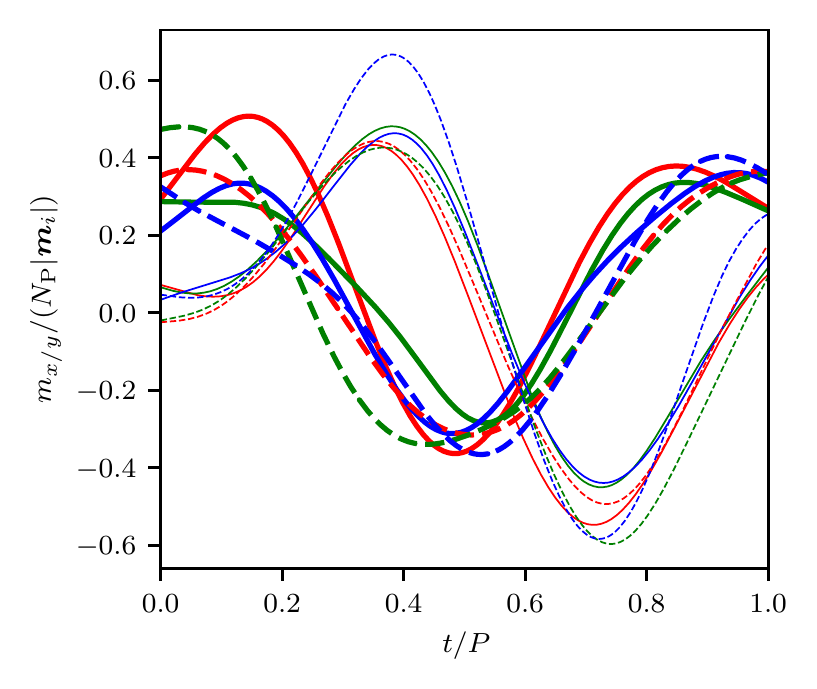}
    \caption{Exemplary trajectories of the total dipole moment in $x$-direction (bold lines) and $y$-direction (thin lines) for a period of $P=32.5\tau_\text{LJ}$. The continuous lines are obtained from the MD simulations, the dashed ones are obtained using interpolated quantities together with the nsGLE.}
    \label{fig:realVsGeneratedInterpTraj}
\end{figure}

Analogously we determine $\mat{\omega}(t)$ and $\left\langle\vec{\eta}_{tt_0}\right\rangle$ by means of interpolation. In \cref{fig:realVsGeneratedInterpTraj} some exemplary trajectories from an MD simulation with $P=32.5\tau_\text{LJ}$ are shown for comparison with the trajectories obtained by propagating the nsGLE with the interpolated memory kernel, drift and noise. The nsGLE with the interpolated coefficient functions produces a bundle of trajectories that have an average and  fluctuations very similar to the MD simulation.

\begin{figure*}
    \centering
    \includegraphics{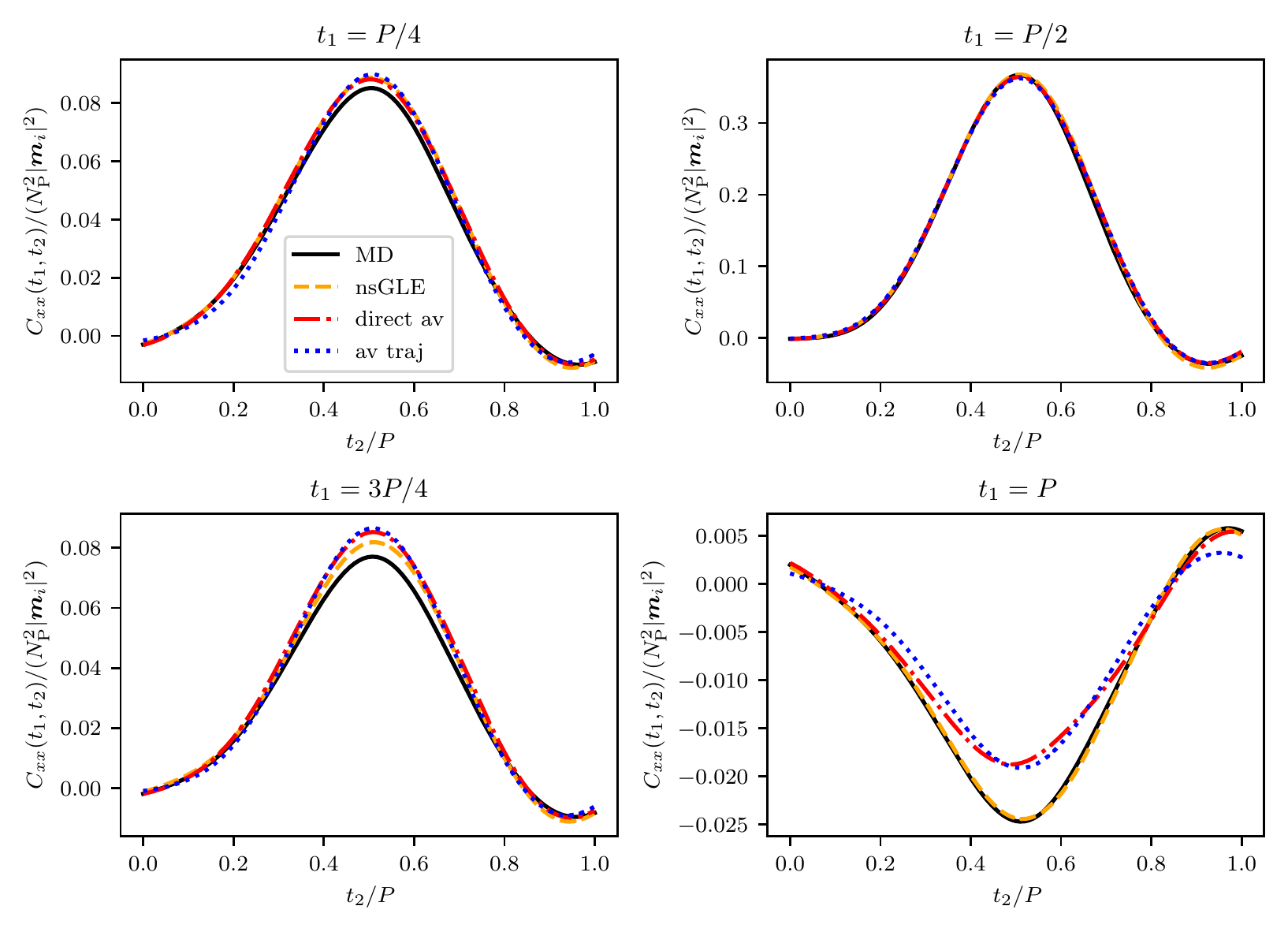}
    \caption{Exemplary slices of the $xx$ component of the two time correlation function for $P=32.5\tau_{\text{LJ}}$. The solid black curve depicts the two time correlation function obtained from running the MD simulations, the dashed orange curve shows the two time correlation function obtained from integrating the nsGLE with averaged quantities. The dash-dotted red line is obtained from averaging the two time correlation functions for $P=30\tau_\text{LJ}$ and $P=35\tau_\text{LJ}$ and the dotted blue line is obtained from averaging the trajectories from the $P=30\tau_\text{LJ}$ and $P=35\tau_\text{LJ}$ simulations.}
    \label{fig:comparisonttCorrs}
\end{figure*}

Finally, we compare the two-time correlation functions of the real trajectories with the one from the trajectories generated with the nsGLE. To check if the interpolation of the coefficient function in the nsGLE has any advantage over directly interpolating either between the correlation functions or between the original trajectories, we also plot the correlation function obtained by an average of the two-time correlation functions for $P=30\tau_\text{LJ}$ and $P=35\tau_\text{LJ}$ (line labeled "direct av" in \cref{fig:comparisonttCorrs})
as well as the two-time correlation function obtained from simply averaging trajectories (labeled "av traj" in \cref{fig:comparisonttCorrs}). By averaging trajectories we refer to the procedure, where one picks randomly one trajectory with $P=30\tau_\text{LJ}$ and one with $P=35\tau_\text{LJ}$ and calculates their mean (again at fixed $t/P$). In \cref{fig:comparisonttCorrs} we see that for $t_1=P/4$ and $t_1=P/2$ there is almost no difference between the interpolation schemes. For $t_1=P$ the results from the interpolated nsGLE  are a much better than the results of the other schemes.  

\section{Conclusion}\label{sec:conclusion}
We have discussed how to replace the fluctuating force of the non-stationary generalized Langevin equation by a stochastic process such that the resulting dynamics reproduces the original, deterministic dynamics on the level of a given set of moments. Combined with a previously published derivation of the non-stationary generalized Langevin equation \cite{MeyerVoigtmann_2019} and with a previously published method to extract the memory kernel from simulation data of the underlying microscopic model \cite{Meyer_Wolf}, the procedure described here allows to construct and to simulate coarse-grained models for systems under time-dependent external driving. 

\section*{Acknowledgment}
The authors acknowledge funding by the Deutsche
Forschungsgemeinschaft (DFG, German Research Foundation) in Projects No. 430195928 and 431945604, and useful remarks from Graziano Amati, Hugues Meyer and Peter Pfaffelhuber.

\section{Data Availability}
The data that support the findings of this study are available from the corresponding author upon request.

%\clearpage
\appendix

\section{Derivation of the generalized Langevin equation}\label{app:gle}
\subsection{Time evolution operator}\label{app:gle1}
In order to derive the time evolution operator, we take the time derivative of eq.~(\ref{eq:time_op}),
\begin{align*}
\partial_t \mathcal{U}(t,t_0) \vec{A}  &= \left[\dot{\Gamma}(t;\gamma(t,t_0,.)) \cdot \partial_\Gamma\right]  \vec{A}(\gamma(t,t_0,.)) \\
&= \mathcal{U}(t,t_0)\left[\dot{\Gamma}(t,.)\cdot \partial_\Gamma\right]  \vec{A} \, .
\end{align*}
Since this equation must hold for arbitrary observables $\vec{A}$, we obtain 
\begin{equation*}
\partial_t \mathcal{U}(t,t_0) = \mathcal{U}(t,t_0)  \mathcal{L}(t) \, ,
\end{equation*}
where we inserted the Liouville operator ${\mathcal{L}(t):=\dot{\Gamma}(t,.)\cdot \partial_\Gamma}$. This equation is solved by a negatively time-ordered exponential, i.e.
\begin{align*}
&\mathcal{U}(t,t_0) =\exp_-\left(\int^t_{t_0}\dd\tau \mathcal{L}(\tau)\right) \, ,\\
&\exp_-\left(\int^t_{t_0}\dd\tau \mathcal{L}(\tau)\right):=1+ \\
& \, +\sum^\infty_{n=1} \int^t_{t_0} \dd\tau_1\int^{t_1}_{t_0} \dd\tau_2 \cdots \int^{t_{n-1}}_{t_0} \dd\tau_n \mathcal{L}(\tau_n)\cdots \mathcal{L}(\tau_1) \, ,
\end{align*}
where $\mathcal{U}(t_0,t_0)=1$.

\subsection{Equations of motion}\label{app:gle2}

The Mori projector satisfies $\mathcal{P}^2(t)=\mathcal{P}(t)$, $\dot{\mathcal{P}}(t)\mathcal{P}(t)=0$ and $\mathcal{P}(t)\mathcal{Q}(t)=0$. With this, the time derivative takes the form
\begin{equation}\label{eq:dot_P}
 \dot{\mathcal{P}}(t) = \mathcal{P}(t)\dot{\mathcal{P}}(t) 
= \mathcal{P}(t)\dot{\mathcal{P}}(t)\mathcal{Q}(t) \, .
\end{equation}

We can now apply eq.~(\ref{eq:dot_P}) to split the dynamics into a projected and orthogonal contribution and use the Dyson-Duhamel identity, as done for the case of a single observable in ref.~\cite{MeyerVoigtmann_2017}. The Dyson-Duhamel identity yields
\begin{align*}
    &z(t,{t_0}):=\mathcal{U}(t,{t_0})\mathcal{Q}(t) \, , \\
	&z(t,{t_0}) = \mathcal{U}(t',{t_0})\mathcal{Q}(t')\mathcal{G}_-(t,t')  + \\
	&\quad +\int^t_{t'} \dd s \, \mathcal{U}(s,{t_0})\mathcal{P}(s)(\mathcal{L}(s)-\dot{\mathcal{P}}(s))\mathcal{Q}(s)\mathcal{G}_-(t,s) \, ,\\
	&\mathcal{G}_-(t,t') := \exp_-\left(\int^t_{t'}\dd\tau \mathcal{L}(\tau)\mathcal{Q}(\tau) \right) \, .
\end{align*}
This representation for $z(t,t_0)$ is easily verified by evaluating its time-derivative and initial value at $t'$. The initial time $t'$ may be chosen arbitrarily which is useful to derive the fluctuation dissipation theorem eq.~(\ref{eq:fdt}).

By inserting this expression into the time derivative of the observable $\vec{A}_t$, we obtain the generalized Langevin equation (nsGLE)
\begin{align}
	\vec{\dot{A}}_t &= \mathcal{U}(t,t_0)\mathcal{L}(t)\vec{A} \\
	&= \mathcal{U}(t,{t_0})\mathcal{P}(t)\mathcal{L}(t)\vec{A}+z(t,{t_0})\mathcal{L}(t)\vec{A} \\
	&=\mat{\omega}(t)\vec{A}_t + \int^t_{t'}\dd s\,\mat{K}(t,s) \vec{A}_s \, +  \mathcal{U}(t',{t_0})\vec{\eta}_{tt'}\, ,  \label{eq:gle} 
\end{align}
where we obtain the final form (\ref{eq:gle2}) for $t'=t_0$. Furthermore, by comparison of eq.~(\ref{eq:gle}) with eq.~(\ref{eq:gle2}), we find
\begin{equation*}
    \mathcal{U}(t',t_0)\vec{\eta}_{tt'} = \int^{t'}_{t_0} ds \, \mat{K}(t,s)\vec{A}_s \,  +\vec{\eta}_{tt_0} \, , 
\end{equation*}
where the time evolution operator on the left shifts some initial value $\Gamma_{t_0}$ at time $t_0$ to the corresponding initial value  $\Gamma_{t'}$ at time $t'$, since both sides of the equation are functions of the initial value $\Gamma_{t_0}$.
The \textit{drift} $\mat{\omega}(t)$ and the \textit{fluctuating forces} $\vec{\eta}_{ts}$ appear in the desired form given by equs.~(\ref{eq:drift})(\ref{eq:ff1})(\ref{eq:ff2}).  The \textit{memory kernel} $\mat{K}(t,s)$ is given by
\begin{align}
	\mat{K}(t,s) &:= \scalar{\left[\mathcal{L}(s)-\dot{\mathcal{P}}(s)\right]\vec{\eta}_{ts}}{\vec{A}}_s\scalar{\vec{A}}{\vec{A}}^{-1}_s \, .\label{eq:definitionMemoryKernel0}
\end{align}
With eq.~(\ref{eq:dot_P}) and $\scalar{\mathcal{Q}(t)\vec{X}}{\vec{A}}_t=\mat{0}$ we find
\begin{align*}
	\scalar{\dot{\mathcal{P}}(s)\vec{X}}{\vec{A}}_s &= \scalar{\mathcal{P}(s)\dot{\mathcal{P}}(s)\mathcal{Q}(s)\vec{X}}{\vec{A}}_s \\
	&= \scalar{\dot{\mathcal{P}}(s)\mathcal{Q}(s)\vec{X}}{\vec{A}}_s \\
	&= \scalar{\mathcal{Q}(s)\vec{X}}{\vec{A}}_s\left[\frac{d}{\dd s}\scalar{\vec{A}}{\vec{A}}_s^{-1}\right] \scalar{\vec{A}}{\vec{A}}_s \\
	 & \quad + \left[\frac{d}{\dd s'}\scalar{\mathcal{Q}(s)\vec{X}}{\vec{A}}_{s'}\right]_{s'=s}\\
	 &=\left[\frac{d}{\dd s'}\scalar{\mathcal{U}(s',r)\mathcal{Q}(s)\vec{X}}{\mathcal{U}(s',r)\vec{A}}_{r}\right]_{s'=s} \\
	&= \scalar{\mathcal{L}(s)\mathcal{Q}(s)\vec{X}}{\vec{A}}_s + \scalar{\mathcal{Q}(s)\vec{X}}{\mathcal{L}(s)\vec{A}}_s \, .
\end{align*}
By applying this identity for $\vec{X}=\vec{\eta}_{ts}$, the memory kernel from eq.~(\ref{eq:definitionMemoryKernel0}) can be written in its final form from eq.~(\ref{eq:definitionMemoryKernel}).

\subsection{Fluctuation dissipation theorem}\label{app:fdt}
With eq.~(\ref{eq:mean_f_A}) and the derivative of eq.~(\ref{eq:gle}) with respect to $t'$
\begin{align*}
    \vec{0} &=-\mat{K}(t,t')\vec{A}_{t'}+\frac{\dd}{\dd t'}\mathcal{U}(t',{t_0})\vec{\eta}_{tt'}\,,
\end{align*}
we find 
\begin{align*}
	&\frac{\dd}{\dd t'}\scalar{\vec{\eta}_{tt'}}{\vec{\eta}_{st'}}_{t'} \\
	&= \frac{\dd}{\dd t'}\scalar{\mathcal{U}(t',{t_0})\vec{\eta}_{tt'}}{\mathcal{U}(t',{t_0})\vec{\eta}_{st'}}_{t_0} \\
	&= \mat{K}(t,t')\scalar{\vec{A}}{\vec{\eta}_{st'}}_{t'}+ 
    \scalar{\vec{\eta}_{tt'}}{\vec{A}}_{t'}\mat{K}^\top(s,t') \\
	&= \mat{0} \, .
\end{align*}
Applying this result to eq.~(\ref{eq:definitionMemoryKernel}) yields the fluctuation-dissipation theorem (\ref{eq:fdt}).

\section{Complex-valued observables}\label{sec:complex_observables}
In principle, the derivation of the nsGLE still holds for complex observables $\vec{A}(\Gamma)\in\mathbb{C}^n$, however, this will lead to a different approach than taking the real and imaginary parts as real observables. We have to choose whether to use the cross-correlation matrix $\langle \vec{X}\vec{Y}^\dagger\rangle$ or pseudo-cross-correlation matrix $\langle \vec{X}\vec{Y}^\top\rangle$ for the definition of the Mori projection operator. Both methods yield valid equations of motion. If we use the cross-correlation matrix, we have $\langle \vec{\eta}_{tt_0} \vec{A}^\dagger_{t_0}\rangle\equiv0$ and the equations of motion for the auto-correlation function $\langle \vec{A}_t\vec{A}^\dagger_s\rangle$, \cref{eq:eom_ac}, still holds. If we use the pseudo-cross-correlation matrix, we have $\langle \vec{\eta}_{tt_0}\vec{A}^\top_{t_0}\rangle \equiv 0$ and \cref{eq:eom_ac} is the equation of motion for the pseudo auto-correlation function $\langle \vec{A}_t\vec{A}^\top_s\rangle$. In either case, one obtains the same statistics up to second order by demanding $\langle \vec{X}'_t \rangle \equiv \langle \vec{X}_t \rangle$, $\langle \vec{X}'_t  \vec{X}'^\top_s \rangle \equiv \langle \vec{X}_t  \vec{X}^\top_s \rangle$ and $\langle \vec{X}'_t  \vec{X}'^\dagger_s \rangle \equiv \langle \vec{X}_t\vec{X}^\dagger_s \rangle$, where the same argumentation as in \cref{sec:second_order} applies. However, it will not be possible to draw the fluctuating forces and initial values independently as proposed in \cref{sec:simplified_procedure}, since in general we have either $\langle \vec{\eta}_{tt_0}\vec{A}^\top_{t_0}\rangle\slashed{\equiv} 0$ or $\langle \vec{\eta}_{tt_0}\vec{A}^\dagger_{t_0}\rangle\slashed{\equiv}0$, depending on the choice of the Mori projection operator. 

More conveniently, we may reduce the dynamics of a complex-valued observable $\vec{O}(\Gamma)\in\mathbb{C}^n$ to the real case, by solving the dynamics for the auxiliary variable $\vec{A}:=(\Re(\vec{O})^\top,\Im(\vec{O})^\top)^\top$. If desired, one may also derive the nsGLE for $\vec{B}:=(\vec{O}^\top,\vec{O}^\dagger)^\top$ by applying the following transformation
\begin{align*}
	\vec{B} &= \mat{M} \vec{A} \, ,  \\
	\mat{M} &= \left(\begin{matrix}
	\mat{\mathbf{1}} & i\cdot \mat{\mathbf{1}}  \\
	\mat{\mathbf{1}} & -i\cdot \mat{\mathbf{1}}   \\
	\end{matrix}\right) \, , \\
	\mat{M}^\dagger \mat{M} &= 2 \cdot\mat{\mathbf{1}} \, .
\end{align*}
If we use the cross-correlation matrix with complex conjugation within its second argument,
\begin{equation*}
 (\vec{X},\vec{Y})_t := \int d\Gamma \, \rho(t,\Gamma) \vec{X}(\Gamma)  \vec{Y}^\dagger(\Gamma) \, ,
\end{equation*}
we recover the same equations of motion for the variable $\vec{B}$ simply by calculating its time derivative. 
\begingroup
\allowdisplaybreaks
\begin{align*}
	\vec{\dot{\vec{B}}}_t &= \mat{M} \vec{\dot{A}}_t \\
	&= \tilde{\mat{\omega}}(t)\vec{B}_t + \mathcal{U}(t',{t_0})\tilde{\vec{\eta}}_{tt'}+\int^t_{t'} \tilde{\mat{K}}(t,s) \vec{B}_s \, ds \, , \\
	\tilde{\vec{\eta}}_{ts} &:= \mat{M} \vec{\eta}_{ts} \\
		&= \mat{M}\mathcal{Q}(s)G_-(t,s)\mathcal{L}(t)\vec{A} \, , \\
		&= \mat{M}\mathcal{Q}(s)\mat{M}^{-1}\mat{M}\mathcal{G}_-(t,s)\mat{M}^{-1}\mathcal{L}(t)\vec{B} \\
		&=\tilde{\mathcal{Q}}(s)\tilde{\mathcal{G}}_-(t,s)\mathcal{L}(t)\vec{B} \, , \\
		\tilde{\mathcal{Q}}(s) &:= \mat{M}\mathcal{Q}(s)\mat{M}^{-1} \\
				&= 1-\tilde{\mathcal{P}}(s) \, ,\\
		\tilde{\mathcal{P}}(s)&:= 	\mat{M}\mathcal{P}(s)\mat{M}^{-1} \\
		&= (.,\vec{B})_s(\vec{B},\vec{B})^{-1}_sB \, , \\
		\tilde{\mathcal{G}}_-(t,s) &:= \mat{M}\mathcal{G}(t,s)\mat{M}^{-1} \\
		&= \exp_-\left(\int^t_s \mathcal{L}(u)\tilde{\mathcal{Q}}(u)du\right) \, , \\
	\tilde{\mat{K}}(t,s) &:= \mat{M}\mat{K}(t,s)\mat{M}^{-1} \\
	&=  -\mat{M}(\vec{\eta}_{tr},\vec{\eta}_{sr})_r\mat{M}^{-1}\mat{M}(\vec{A},\vec{A})^{-1}_s\mat{M}^{-1}  \\
	&= -(\tilde{\vec{\eta}}_{tr},\tilde{\vec{\eta}}_{sr})_r(\vec{B},\vec{B})^{-1}_s \, ,\\
	\tilde{\mat{\omega}}(t) &:= \mat{M}\mat{\omega}(t)\mat{M}^{-1} \\
	&= \mat{M}(\mathcal{L}(t)\vec{A},\vec{A})_t\mat{M}^{-1}\mat{M}(\vec{A},\vec{A})^{-1}_t \mat{M}^{-1} \\
	&= (\mathcal{L}(t)\vec{B},\vec{B})_t(\vec{B},\vec{B})^{-1}_t \, .
\end{align*}
\endgroup
Hence, for complex-valued observables $\vec{O}$, we may simply define $\vec{A}:=(\vec{O}^\top,\vec{O}^\dagger)^\top$ and continue as usual, if we consequently use the cross-correlation matrix with complex conjugation in its second argument. Note that the corresponding auto-correlation function $\langle \vec{A}_t\vec{A}^\dagger_s\rangle$ solves eq.~(\ref{eq:eom_ac}) and determines both, $\langle \vec{O}_t\vec{O}^\dagger_s\rangle$ as well as $\langle \vec{O}_t\vec{O}^\top_s \rangle$, as needed for second order statistics. Further, we have $\langle \vec{\eta}_{tt_0}\vec{A}^\dagger_{t_0}\rangle\equiv\langle \vec{\eta}_{tt_0}\vec{A}^\top_{t_0}\rangle \equiv 0$ and all results from sec.~(\ref{sec:numerical_method}) apply without restrictions.

\section{Notation}
\begin{itemize}
\item $\equiv$: Equality for all times $t,s$ whereas $t_0$ remains arbitrary but fixed.
\item $\dagger$: Conjugate transpose.
\item $\langle X \rangle :=\mathbb{E}[X] $: Expected value of some random variable $X$.
\item $\mat{\mathbf{1}}$: Identity matrix.
\item $\otimes$: Tensor product.
\item $\mathcal{O}$: Calligraphic symbols denote operators.
\item $\vec{X}$: Bold symbols denote vector-valued objects.
\item $\mat{M}$: Underlined symbols are matrices.
\item $\mmat{M}$: Block matrices.
\item $F_t$: Time as an index indicates that $F_t$ is a function of the phase space variable $\Gamma_{t_0}$ and not an ensemble-averaged quantity.
\end{itemize}

\end{document}